# Classification of cyber-physical production systems applications: proposition of an analysis framework

Olivier CARDIN

LUNAM Université, IUT de Nantes – Université de Nantes, LS2N UMR CNRS 6004
(Laboratoire des Sciences du Numérique de Nantes),
2 avenue du Prof. Jean Rouxel – 44475 Carquefou
(e-mail: olivier.cardin@ls2n.fr).

**Abstract.–** Cyber-physical systems have encountered a huge success in the past decade in several scientific communities, and specifically in production topics. The main attraction of the concept relies in the fact that it encompasses many scientific topics that were distinct before. The downside is the lack of readability of the current developments about cyber-physical production systems (CPPS). Indeed, the large scientific area of CPPS makes it difficult to identify clearly and rapidly, in the various applications that were made of CPPS, what are the choices, best practices and methodology that are suggested and that could be used for a new application. This work intends to introduce an analysis framework able to classify those developments. An extensive study of literature enabled to extract the major criteria that are to be used in the framework, namely: Development Extent; Research Axis; Instrumenting; Communication standards; Intelligence deposit; Cognition level; Human factor. Several recent examples of CPPS developments in literature are used to illustrate the use of the framework and brief conclusions are drawn from the comparative analysis of those examples.

**Keywords:** Cyber-physical production systems, classification, cognition, framework, human factor, Manufacturing control.

## 1  Introduction

First definition that can be found about cyber-physical systems (CPS) dates from 2006 [1], during a workshop with the American National Science Foundation[1] (NSF). The extension of cybernetic systems towards CPS is therefore explicitly dealt with in literature since 2006-2007 and is constantly growing in popularity. Fig. 1 shows a short analysis of the evolution of the number of journal article mentioning explicitly the term "cyber-physical" in the 5 major scientific publishers. This term was chosen as it encompasses various aspects and applications of the CPS. In the past decades, many research articles were dealing about notions and concepts that have been at the origin of current CPS. These works are not listed here as they do not mention the cyber-physical keyword even though the scientific content is compatible. The objective of this figure is to exhibit the trend of acceptance of the notion in literature, whatever the field of research and the publisher. Over the last six years, i.e. since the total number of publications reached 1000, a 40% increase in average per year can be noticed, which demonstrates the high level of acceptance of the notion. Looking at the publisher proportions, it can be noticed that IEEE was very present in the early years, whereas the publication rate is more shared nowadays. This is an effect of the dissemination of the notion to a large number of new fields of application that were not present in the first years.

---

[1] http://www.nsf.gov/

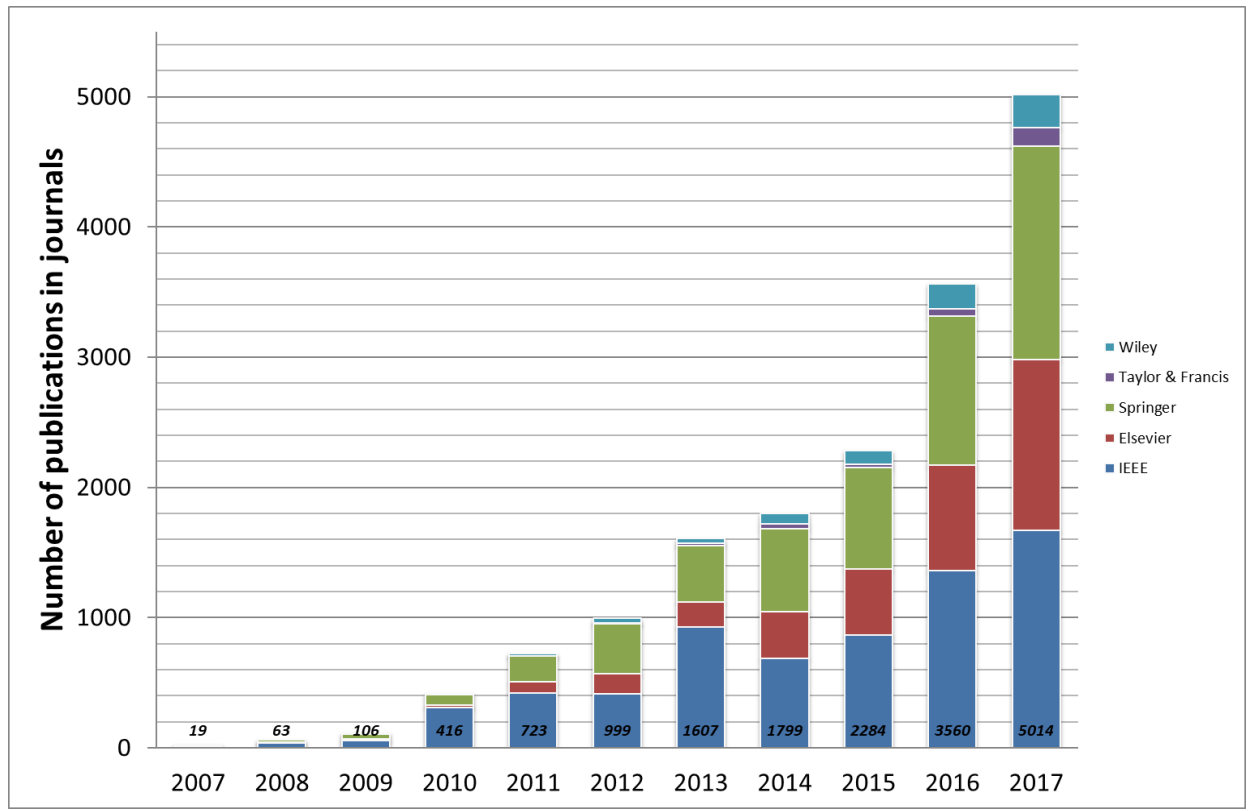

**Fig. 1.** CPS notion dissemination in literature

All along their development, more synthetic definitions were suggested, such as those of [2] or [3]. For example, CPS are defined by [4] as cooperating systems, having a decentralized control, resulting from the fusion between the real world and the virtual world, having autonomous behaviors and dependent on the context in which they are, being able to constitute in systems of systems with other CPS and leading a deep collaboration with the human. For this, embedded software in CPS uses sensors and actuators, connect to each other and to human operators by communicating via interfaces, and have storage and data processing capabilities from the sensors or the network [5]. The recent one, suggested by [6], allows a clear synthesis of the various aspects of this large concept, coupling in addition the notion of services with CPS : *"Cyber-Physical Systems (CPS) are systems of collaborating computational entities which are in intensive connection with the surrounding physical world and its on-going processes, providing and using, at the same time, data-accessing and data-processing services available on the internet"*. To do so, embedded software in CPS uses sensors and actuators, connect with each other and with humans communicating via standard interfaces, and have abilities of storage and processing of data coming from sensors or from the network [5]. This interconnection of systems, as stated by [7], derives from the fact that a CPS encompasses together control, computation but also communication devices [8]. What can be spotted in the evolution of the definition is the notion of system of systems that was not considered in the early definition of 2006 [1].

The notion of CPS is very wide and encompasses an extremely large class of systems. As a matter of fact, numerous fields of research are relevant of this keyword. This is probably a huge opportunity as it gives the possibility to create a consistent ecosystem in numerous fields of applications: from autonomous vehicles [9] to health devices [10], from electrical grid management [11] to HVAC building control [12].

The industrial domain is of course interested in this field, and the development of major evolutions such as Industrie4.0 in Germany [13] is based on CPS. The application of CPS in the field of production management was formalized in the past few years [6], under the term of Cyber-Physical Production Systems (CPPS). According to [14], the main benefits that can be expected from the generalization of CPPS are: (i) Optimization of production processes; (ii) Optimized product customization; (iii) Resource-efficient production; (iv) Human-centered production processes.

Many fundamental research questions emerge from the development of the concept of CPPS. Among these, the modelling and forecasting of their emergent behavior, the optimization of the control laws at each level of the system and the establishment of a convenient environment for developing autonomy, cooperation, optimization and responsiveness can be cited [6,15]. In parallel, a credible answer to this last element was given thanks to the development of Holonic Manufacturing Systems [16–20], but still needs to be developed in order to fit the requirements of industrial implementation at a large scale. On the other hand, a special interest is being given nowadays to cloud technologies, which are becoming a more and more credible actor of future industrial systems [21–23]. Projects such as IMC-AESOP [24] for example studied the benefits that can be expected from the coupling between cloud and CPPS, and technologies such as Service-oriented Architectures (SoA) are given a certain credit in order to foster interoperability, agility and self-* abilities of systems [25]. The human-machine interaction [26,27], the social aspect [28] and the cyber-security issues [29–31] applied to industry and manufacturing are also major questions that are under study and directly connected to CPPS.

Considering this extremely large field of research and the large perimeter of the definition of CPPS, there is a risk of scattering of research efforts inside this wide notion. The main challenge of the next few years is to provide proofs of concepts, industrial applications and laboratory developments able to prove the advantages given by the CPPS paradigm in terms of flexibility and performance. This article intends to introduce an analysis framework aiming at classifying the various developments and applications of literature. This framework is intended for future researchers or engineers willing to establish rapidly an overview of the main trends of literature developments.

First, an analysis of the research context from the international roadmaps perspective is presented. Then, the framework will be presented and the items defining the axes of the framework are described. Finally, the application of the framework to various examples of developments found in literature in order to illustrate the use of the framework is introduced, and a preliminary analysis of the state of the art based on the use of the framework is proposed in the agility domain.

## 2 Cyber-physical production systems research context

### 2.1 Definition and fundamentals

CPPS classical definition [6] is widely accepted in the last few years as it exhibits well the notion of the necessary cooperation between CPS in a CPPS. However, the notion of knowledge management and decision making, which constitute still nowadays a large field of research, are missing. Notions such as digital twins used for dynamic simulation and forecasting are not present for example. Furthermore, notions of learning and auto-adaptation are not clearly mentioned. Finally, the adaptability of the system to new technologies, new organizations or major reconfigurations are not expressed. As a matter fact, clearly defining the goals, objectives and benefits obtained by a paradigm shift towards CPPS is difficult to establish.

In this article, the following definition, adapted from [6], is suggested in order to encompass these missing notions: *"Cyber-Physical Production Systems are systems of systems of autonomous and cooperative elements connecting with each other in situation dependent ways, on and across all levels of production, from processes through machines up to production and logistics networks, enhancing decision-making processes in real-time, response to unforeseen conditions and evolution along time"*.

As this definition shows, if the notion of CPPS is new and brings many different fields of research together towards high-level objectives, the fundamentals of the notion considered one by one are not really new. Several flagship notions in the past decades were already addressing the same objectives with globally the same ideas of solutions, among which Intelligent Manufacturing Systems [32], Biological Manufacturing Systems [33], Reconfigurable Manufacturing Systems [34], Digital Factory [35], Holonic Manufacturing Systems [36], Industrial Agents [37], etc. However, the major benefit of embracing CPPS is that it is able in the next few years to be accepted by the largest audience and gather all the publications in a single trend, that shall make the readability of the scientific community more clear and thus more visible.

If CPS characterization in Computation, Control and Communication [8] is currently widely disseminated, CPPS were also characterized many times in various ways by the different authors of literature. An interesting synthesis of these various characterizations is however defined in [38], which states that CPPS are globally characterized by three main characteristics: Intelligence, Connectedness and Responsiveness. Fig. 2, adapted from [8], positions the main characteristics of CPS and integrates the fundamentals of CPPS. It exhibits how all these characteristics finally fit well together and shows the clear relation between these notions, each in their field. In the following, the basic characteristics of the CPS will therefore be extended to CPPS for clarity purposes.

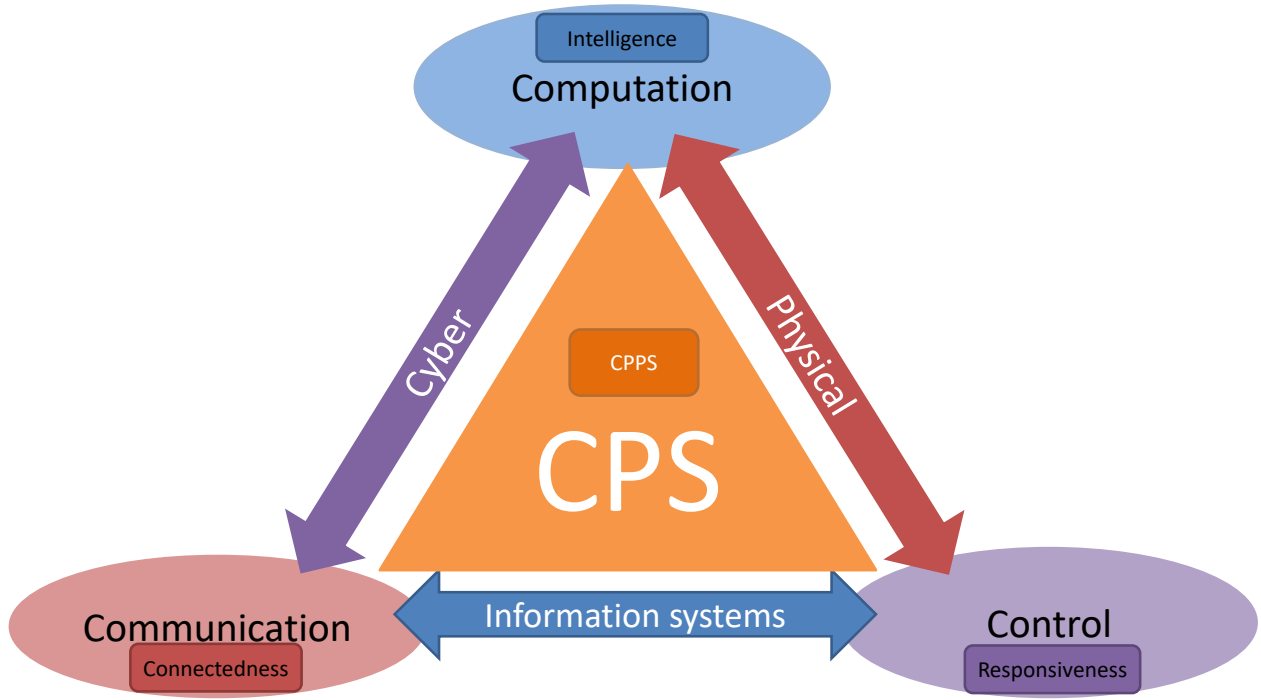

**Fig. 2.** Basic capabilities of a CPS and the analogy to CPPS

### 2.2 Research context overview

In order to establish the context in which the researches in CPPS field are developed, the idea here was to concentrate on the international research roadmaps in order to exhibit the main trends in the field. A study of these roadmaps in literature [39], classifies these international roadmaps towards the main axes they put forward. Several conclusions can be drawn from this comparative study. First, it may be noted that although different terminologies are used, these different roadmaps provide a fairly coherent and identical view of both the objectives and the means of achieving these objectives. Moreover, innovation objectives can be classified into three major development axes, relatively stable regardless of the roadmap studied over the last decade.

The first axis is the development of advanced technologies for production. The idea of offering high technology in manufacturing processes is assumed to have in the near future an important impact on next-generation production systems. Current trends focus on nanoscale structures, additive manufacturing (e.g. the PHOCAM-Photopolymer-based Customized Additive Manufacturing Technologies project) or intelligent materials (e.g. project AMITERM[40]). All these processes have the particularity of being more and more connected, which implies notions of cyber-security that are present transversally in many projects.

The second axis deals with sustainable production. In order to reduce the impact of production on the environment, an initial idea is to increase the energy efficiency of processes, as in projects such as Co2PE! (Cooperative effort on Process Emissions in manufacturing)[41,42], DAPhNE (Development of Adaptive ProductioN systems for Eco-efficient firing processes), EEM (Energy Efficient Manufacturing) or Factory Ecomation (Factory ECO-Friendly and energy efficient technologies and adaptive automation solutions). A second idea is to develop closed-loop (PLM) lifecycle concepts in the context of a dynamic supply chain (e.g. PROMISE [43]). The objective is to be able to identify, retrieve and reuse useful components of obsolete or dysfunctional equipment in order to reduce the environmental impact of the design and production of new equipment (e.g. SuPLIGHT project: Sustainable and Efficient Production of Lightweight Solutions). The place of humans in the new workshops is also studied, because it is necessary to design the future environments in which human operators and new technologies can evolve in total cooperation. The aim is to create human-centered production systems, as is clearly illustrated by the SO-PC-PRO (Subject-Orientation for People-Centered Production) project, to study the respective layouts of each element , with, for example, MAN-MADE (Manufacturing through ergonoMic and safe Anthropocentric aDaptive workplaces for context aware factories in Europe), and the whole virtual operator training phase, with VISTRA (Virtual Simulation and Training of Assembly and Service Processes in Digital Factories ).

Finally, the third axis of development aims at the agility of systems, i.e. collaborative, mobile, intelligent and adaptive systems. They are supposed to be able to change, reconfigure and evolve rapidly over time, depending on market demand. A lot of work is dedicated to the interoperability of these systems, in an open production context (e.g.

ACMN - Automation Competency Model Network or LinkedDesign - Linked Knowledge in Manufacturing, Engineering and Design for Next Generation Production), or to mass customization, from the workshops to the supply chain as for example in MIGOODS (Manufacturing Intelligence for Consumer Goods: Fit4U, SShoes, A-Footprint). Finally, the need for reconfigurable and flexible systems is expressed through concepts such as distributed intelligence, industrial agents or intelligent products. Projects such as ARUM [37,44], PABADIS [45], GRACE [46] or ERRIC [47] have shown great advances in the introduction of distributed / holonic / multi-agent technologies and paradigms in manufacturing control.

## 3    Framework items description

The core of the framework is to design a set of characteristics that are of interest for classifying the applications and implementations of CPPS in literature. To do so, one possibility was to analyze the main characteristics and objectives of global CPPS and deduce the criteria of the framework. Exploring this possibility resulted in various criteria that were not directly related to the developments of the CPPS, but rather on the methodology and objectives.

In this work, the different points of view of a CPPS application were analyzed, and some corresponding criteria were expressed. To do so, the three main basic capabilities of a CPS expressed by [8] were adapted, namely Communication, Computation and Control (see Fig. 2) in addition to a global point of view on the application, as CPPS clearly inherit from CPS and therefore can be expressed through the same scope of basic capabilities:

- CPPS in global:
    - What is the degree of maturity of the considered CPPS application?
        - Criteria 1: development extent
    - What is the main objective of the considered CPPS application?
        - Criteria 2: Scientific research axis of development
- CPPS Communication:
    - How is the CPPS information system connected? Literature can expect a large evolution of this criteria considering the development of IoT, fog computing and high speed networks;
        - Criteria 3: Communication standards
- CPPS Computation:
    - As the CPPS concept is oriented towards decentralized control, where is the intelligence deposit in the considered CPPS application?
        - Criteria 4: Intelligence deposit
    - What is the maturity of the knowledge management issue of the considered CPPS? This criterion constitutes an alternate way to evaluate the level of intelligence of the CPPS without addressing the notions of performance, which are generally very specific to the studied system;
        - Criteria 5: Cognitive abilities
- CPPS Control:
    - Considering the CPPS in itself, how is the CPPS sensing its physical environment? This criteria encompasses auto-id technologies for example;
        - Criteria 6: Pervasive instrumenting
    - How is the CPPS interacting with humans in its environment and control?
        - Criteria 7: human-machine interface.

Fig. 3 represents a graphical view of the framework and shows how it inherits from the decomposition adapted from [8] and presented before.

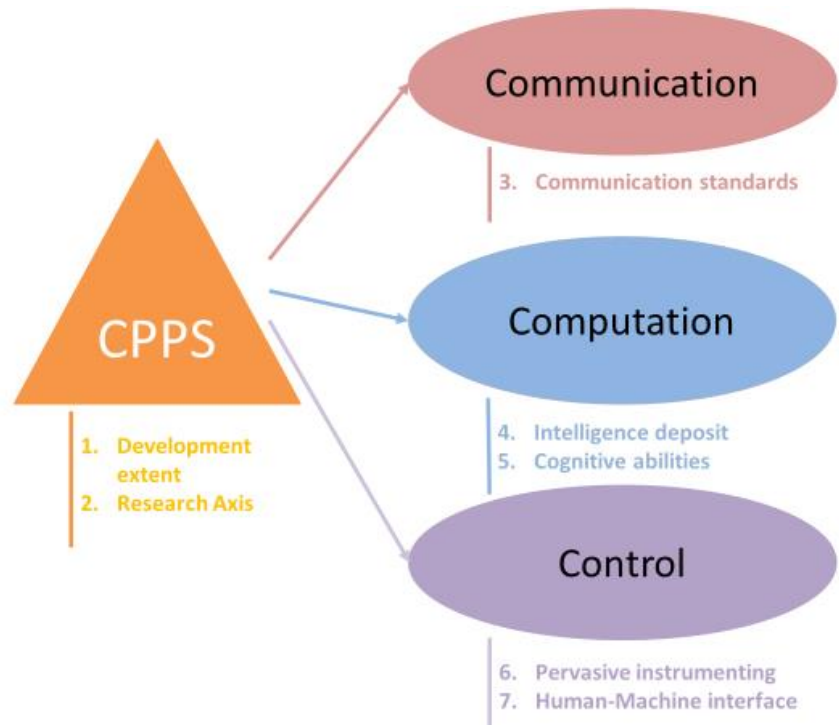

**Fig. 3.** Graphical expression of the framework

The following subsections detail these criteria.

### 3.1 Development extent

The first item to specify the CPPS developments is designed to indicate the extent to which the development was made. To this end, four different levels were specified:
- Lab XP: this level deals with Laboratory eXPeriments where technologies are developed and evaluated. This extent can be found currently for co-manipulation of robots, connectivity of CNC or development of new communication standards for PLCs, where the notion of collaboration is only evaluated at a quasi-single level;
- POC: the Proof Of Concept level is dedicated for laboratory setups including the notion of collaboration between CPPS and a behavior close to the one that could be found in an industrial context. This level is currently representative of the large-scale demonstrators that can be found in [48–52] for example;
- Industry: the objective of these applications is to reach the industrial application level, where actual developments are made on a running production system. At this level, coherent performance evaluation can be made and feasibility purposes can be identified;
- Learning factory: one of the biggest challenge that will face industry when CPPS-based manufacturing systems will be extensively available will be to have engineers and technicians trained to work in full cooperation with those kind of systems. This is why, as soon as possible, Learning Factories need to be developed in order to foresee the arrival of CPPS technologies on the market.

### 3.2 Research axis

Considering the analysis that was developed in section 2, this item is meant to indicate what is the aspect of CPPS evolution the development emphasizes the most. As suggested and analyzed in [39], the considered level for this item for the time being are:
- Agility;
- Technology;
- Sustainability.

However, this list is meant to evolve in the next few years with the resolution of some of the issues and the creation of new research items. For example, it is clear that the current development are specifically focused on providing additional functionalities to CPPS comparing to traditional manufacturing systems. However, in the next few years, economic questions about cost (implementation and exploitation) and productivity of these systems will probably occur.

### 3.3 Communication standards

The mutation in the high-level control architecture of production systems brings about potentially significant changes in the technology of the production systems themselves. The control of production machines is currently at a very low semantic level, often of the order of bit or word, via communication protocols inherited from industrial local networks developed in the last 40 years.

From our perspective, the constraints requiring the use of such protocols are not justified throughout the workshop. Indeed, their first quality is the determinism of their behavior, which imposes constraints on the volume of data exchanged so as not to saturate the network. However, while some functions require this determinism and a certain notion of real-time within the control, other coordination functions do not have this need: this aspect is often referred to as real-real-time (near real- time) in the literature. While some cloud-based work proposes PLC-free control architectures, we propose rather a technological breakdown according to the objectives and constraints of each part of the system. Figure 4 provides an example of an instantiation of the SoHMS model deployed in the cloud presented in [53]. It is very clear that robots always need a controller, just as it seems unreasonable to completely remove controllers as close as possible to sensors and actuators in the case of high-speed production systems. Moreover, it is possible to consider that some systems, controlled by separate systems (two systems, each with an API, for example) need a low-level communication medium as described above. The level of atomic aggregation of the CPPS is generally considered as the element of the system which cannot be dissociated without altering its communication performance with the other CPPS. In the case presented earlier, the proposed CPPS would encompass both systems as if they were one. Thus, the communication between CPPS would be only in soft real-time, which makes it possible to envisage communications at higher semantic level, in particular by using Service-oriented Architectures [54].

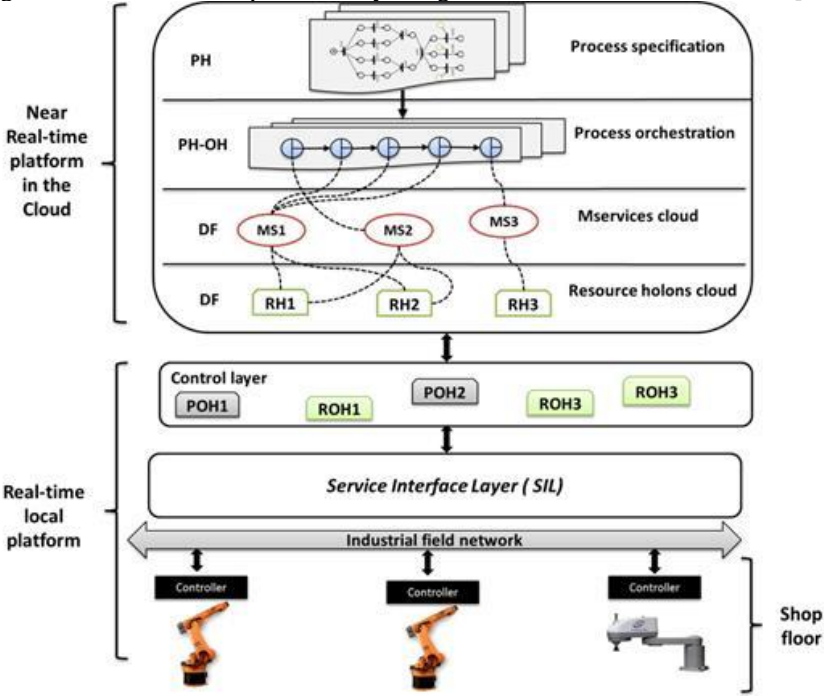

Figure 4 Example of instantiating the SoHMS architecture in a cloud context [22]

The major development axis still to be studied at the technological level corresponds to the SIL layer presented in Figure 4. Through this architecture, the transition between the services resulting from the cloud and the communication protocol accepted by the low-level entities must be realized. The aim is to propose an open protocol allowing to minimize or even eliminate this transition by increasing the semantic level accepted by the entities at their superior interface. It is therefore necessary to develop PLCs, robot controllers or service-oriented CNC machines, not only in their communication, but also in the parameterization of the programs executed on these elements. Indeed, at the programming level, new elements have to be proposed in order to simplify the modularity of the programs and their reusability. At the communication level, attempts at solutions have already been proposed, proprietary such as OPC-

UA [55] or not such as MSB (Manufacturing Service Bus) [56], but their integration with the concept of services remains to be studied and developed.

The current item aims at indicating which communication protocols are used:
- If available in the lower level, close to the machines;
- If available in the higher level, close to the cloud or enterprise information systems;
- If necessary, the protocols matching the communication between the lower and the higher levels.

### 3.4 Intelligence deposit

The notion of CPPS encompasses a wide variety of control architectures, having the common feature to avoid the rigid hierarchy that can be currently encountered very frequently. In those so-called heterarchic architectures [57], the notion of intelligence is widely distributed within the whole system, even if a centralized coordination is still present. This presence can dynamically evolve throughout the life of the system, considering the disturbances occurring during the production [58].

Nevertheless, this classification item is meant to represent the location where the intelligence of the system is distributed. The most common levels this item can have are:
- Machines: the most common decentralization of intelligence currently found in literature deals with the enhancement of machines' abilities in terms of decision making. The machines are for example able to negotiate and establish their own planning in order to decide whether they accept the contract to be established by an order [59];
- Products: many works in the last decade were focused on the activity of the product all along the production phase [60–62]. Meanwhile, several parallel directions lead to the concept of "Intelligent Product" [63–67], where the product was the trigger and a true actor of the all the decisions that were dynamically taken during the production thanks to storing and processing capabilities that are physically implemented on the product himself. This concept is now globally replaced by a more cooperative one, where the product can still be a carrier of information or of some distributed intelligence, but is rarely alone in the system.
- Transport: In many applications, part of the intelligence is distributed among the resources enabling the products to be conveyed from one machine to another, may it be using AGVs [68] or conveyors [52,69] for example;
- Digital twin: For some applications, the intelligence is not physically implemented in the machines, products or any other assets of the system, but is directly connected to a digital twin of these entities [70]. This aspect of intelligence is especially relevant when the products cannot be physically equipped with network adaptor, data carriers or data process units or if the legacy systems that are actually used are not yet able to implement those elements.

### 3.5 Cognition level

Such abilities are often referred as self-adaptation, self-organization or more generally self-* abilities [71]. The objective of this criterion is to evaluate the degree of maturity of the development regarding cognition. A classification of CPS in five categories (5C) according to the level of integration it offers was presented in [72] and can be extended to CPPS. This classification was chosen for the pertinent graduation it offers, and is getting more and more attention from the community [73]. It can be described in the following way:

- **C1.** At Connection level, CPPS operate on a Plug&Play network and use data provided by sensors on the network;
- **C2.** At Conversion level, CPPS process data and aggregate them in a higher semantic level;
- **C3.** At Cyber level, CPPS can apprehend other CPPS and their environment and can interact with them in order to enrich their own data processing;
- **C4.** At Cognition level, CPPS are able to process data in order to diagnose their own state, based on simulations and a differential analysis of sensors data;
- **C5.** At Configuration level, CPPS shall adapt on their own facing disturbances, reconfigure or adjust their parameters in an autonomous way in order to get back to a nominal behavior as soon as possible.

This classification was initially introduced in order to provide a step by step CPPS deployment tutorial, from sensing functionalities to functions creating more added value. The total integration of these 5 levels in a CPPS is currently extremely rare and might not be pertinent in any situations. We suggest in this article to extend the use of this functional decomposition and use it in order to evaluate the level of autonomy and intelligence embedded in a given CPPS. This constitutes the cognition-oriented criterion of the framework that we suggest to define (C1 to C5).

### 3.6 Pervasive Instrumenting

One of the core notions of CPS, and therefore of CPPS, is the notion of connectivity to its environment. Indeed, those systems are meant to be able, at a high degree of changeability, to sense the evolution of their direct environment and make their own behavior evolve in consequence, by themselves or via a negotiation with the other elements of the CPPS. This is one of the main notions that make the design of CPPS different and difficult to apprehend.

Even if this is not quite yet fully implemented in CPPS, it is necessary to anticipate the next step of CPPS development, which will be able to take into account more numerous and more complex data, including data coming from various sensors and sources in its environment. Currently, RFID tags and readers, enabling the identification of the moving elements throughout the system are the main equipment that are implemented in the large applications [74]. For smaller ones, such as Lab XP for example, some sensors start to be implemented benefiting from the advances in the field of the Internet of Things for example [75].

### 3.7 HMI: Human-Machine Interaction

The objectives of systems integration evolved in the last few years, shifting from a model where the system was intended to adapt automatically and be equipped with reasoning and decision making capabilities aiming at replacing human ones to a model where CPS are focusing on use scenarios fully involving the humans. Two possible scenarios [76] were specifically designed.

In the first one (called Automation scenario), the human is guided by the CPS, i.e. the global decision making process is performed by the CPS and the human executes the operation in itself. It is also a human that is responsible of the implementation and maintenance of the CPS. In a manufacturing context, his scenario corresponds roughly to an online and pulled flow oriented transposition of the classical functions of planning and scheduling that can be encountered currently. This scenario is intended to fit well to the workshops implying a heavy manual duty in a manufacturing environment targeting a high flexibility.

Second scenario (Tool scenario) emphasizes the human in the core of the decision loop. The idea is to make the CPS being guided by a human initiated to the cooperation with the CPS, but still actively assisting the human in the decision making process. This scenario intends to fit the activities in which operations are partially or fully automated, but where the operator's expertise brings a significant contribution notably in terms of agility and quality enhancement.

In both first and second scenario, it is the combination between calculation abilities of CPS and communication with human capacities that enable the enhancement of the performance of the cooperation system. Frameworks such as HilCP²SC (Human-in-the-Loop Cyber-Physical Production Systems Control) are being developed [77], in order to offer the possibility to integrate the preferences of the human in a multi-objective decision making context led by the CPS. More than a framework, it is probably necessary to modify the (manufacturing) distributed systems design paradigm, using CPS reference models that need to be anthropocentric, such as those suggested by [78] or [79] for example. Whatever the scenario, a significant evolution of the tasks, qualifications and skills of the human operators in charge of the cooperation with the CPS [80] is foreseen. Therefore, basic and professional trainings of those operators needs to adapt in order to give them the keys to efficiency in a problem resolution process, to have a more accurate conscience of process interdependencies of which they are a link and to be able to take regulated initiatives for self-organization in case of disruptions occurring in the nominal working of the system.

As it can be noticed, the CPS-Human integration is of a great matter, and considers a whole scope of situations, from the most automated ones to the most manual ones. In production systems, The same scope of situations exist, and can be characterized in the same way, although some of them are not so much disseminated for the time being. Therefore, considering this criterion, a larger list was induced from the levels suggested by [76] and applied to CPPS:

- **Full** : the human only has a role of supervision of the CPPS, which is able to take all the necessary decisions without any intervention of the human;
- **Automation**: the CPPS guides the human during its task by taking most of the decisions and leaves the functions of adaptation to the human;
- **Tool**: the human guides the CPPS and is in charge of most of the decisions;
- **Manual**: CPPS only provides data to the human, who is in charge of all the decisions.

# 4 Example of application of the framework to a subset of current literature

The objective of this article is to introduce an innovative framework enabling the classification of cyber-physical production systems developments in literature in order to identify either the convergence between the different authors or the leads that were not explored for example. In this section, a short example of use of the framework and a preliminary analysis is performed in order to illustrate the benefit of such a framework for future potential users.

The focus of this analysis was chosen on the Agility axis. As stated before, the concept of CPPS inherits from many older fields of research, among which Intelligent Manufacturing Systems, Reconfigurable Manufacturing Systems, Holonic Manufacturing Systems, Industrial Agents. Most of these fields are targeting the Agility axis, which explains why many of the early developments of CPPS available in literature are related to this axis. This analysis was oriented towards identifying the relationship between the extent of the developments and both the cognition level of the proposed solution, which is closely related to its intelligence, and the interaction with humans.

## 4.1 Analysis grid of current literature regarding Agility

The articles exposed in this section are representative of the current developments currently available in literature and specifically addressing the concept of CPPS, restricted to those aiming for the Agility axis. It is obvious that a lot more applications can be encountered without clearly addressing the notion of CPPS, but these are not included in the limits of this analysis. **Table 1** positions some of the most recent articles in the field relatively to all the items expressed before. The columns represent in the same order the items described in the previous section contained in the framework.

**Table 1.** Agility oriented application of the framework example

| Reference | Extent | Axis | Communication | Intelligence | Cognition | Instrumenting | HMI |
|---|---|---|---|---|---|---|---|
| [23] | Lab XP | Agility | TCP/IP & USB | Machines | C2 | Sensors | Full |
| [26] | Lab XP | Agility | TCP/IP | Machine | C2 | Sensors | Tool |
| [81] | Learning Factory | Agility | RFID | Products and machines | C1 | Sensors | Tool |
| [82] | Lab XP | Agility | TCP/IP | Machines | C2 | Unknown | Tool |
| [83] | Lab XP | Agility | TCP/IP | Products digital twin | C2 | Unknown | Full |
| [84] | Learning Factory | Agility | TCP/IP | Products and machines | C3 | RFID & sensors | Automation |
| [85] | Lab XP | Agility | PROFINET | Central | C1 | Sensors | Full |
| [86] | POC | Agility | TCP/IP | Machines | C2 | Sensors | Full |
| [87] | Learning Factory | Agility | TCP/IP | Central | C1 | RFID | Manual |
| [88] | POC | Agility | TCP/IP | Machines and Transport | C2 | Sensors | Tool |
| [89] | Lab XP | Agility | TCP/IP | Machines | C2 | RFID & Sensors | Tool |
| [90] | Lab XP | Agility | TCP/IP | Machines | C1 | Sensors | Tool |
| [91] | Lab XP | Agility | TCP/IP | Digital Twin | C2 | Sensors | Full |
| [38] | Lab XP | Agility | OPC UA | Central | C1 | Sensors | Full |
| [92] | Lab XP | Agility | RS232 | Digital Twin | C1 | Sensors | Full |
| [93] | POC | Agility | Wifi | Machine | C2 | Sensors | Tool |
| [11] | Lab XP | Agility | TCP/IP | Machine | C2 | Sensors | Full |
| [94] | Industry | Agility | Wifi | Machine | C3 | Sensors | Tool |
| [95] | POC | Agility | Ethernet | Machine | C2 | RFID | Tool |
| [96] | POC | Agility | Ethernet | Machine | C2 | RFID | Tool |

| [95] | POC | Agility | Ethernet | Product | C2 | RFID | Tool |
| [97] | POC | Agility | OPC UA | Machine | C2 | Sensors | Tool |
| [98] | Lab XP | Agility | TCP/IP | Machine | C2 | Sensors | Tool |
| [99] | Lab XP | Agility | OPC | Machine | C2 | RFID | Tool |

The data from Table 1 can also be represented with a literal expression of the framework, especially useful in text paragraphs for example:

*Reference = {Extent; Axis; Instrumenting; Communication; Intelligence; Cognition; HMI}*

With this notation, some examples can be given from the **Table 1**:

[86]= *{POC; Agility; TCP/IP; Machine; C2; Sensors; Full}*

[91]= *{Lab XP; Agility; TCP/IP; Digital Twin; C2; Sensors;Full}*

[94]= *{Industry; Agility; Wifi; Machine; C3; Sensors;Tool}*

### 4.2 Preliminary remarks regarding Instrumenting and Communication protocols

Looking at Table 1, first elements that can be emphasized concern the characteristics with an open choice. Instrumentation devices and communication protocols are indeed connected to technological aspects, that cannot be listed exhaustively at a given time.

Table 1 exhibits some values of those criteria that are not really consistent. Indeed, considering the communication protocols for example, some of them belong to the same class, while others can run on top of the rest. This apparent confusion in the list is mainly due to the lack of information that is present in the examined articles. This lack of information is a true handicap for the readers who are potentially interested in implementing its own CPPS, as this is an important matter. The same reasoning fits to the Instrumenting criterion.

The framework introduced above aims at emphasizing these criteria in order to encourage the publication of such data and disseminate the good practices in literature.

### 4.3 Using the framework to draw preliminary conclusions on CPPS developments targeting Agility

First conclusion that can be drawn from this table is the high density of applications between C1 and C3 (see also Table 2). Globally, it was expected to encounter relatively few applications with C5 abilities, as these cover at the time being more perspectives than actual possibilities. However, the low presence of C4 levels is surprising in manufacturing comparing to fields of application like Infrastructure [12] or Transportation [100] for example. This can be explained by the late appropriation of the concept by manufacturing, and the high inertia of the field of application due to the current limitations of high cost hardware that is not suitable for the full integration of high cognition levels.

Most of the listed applications are currently Lab XP, which is coherent with the global state of the art. However, some POC and industrial applications are rising, which is an interesting indicator for the future developments of literature. The important presence of Learning factories is interesting, as it shows a great interest of the community towards this difficult question of training.

Crossing these criteria in Table 2 exhibits a major axis of research that is still not explored considering levels C3, C4 and C5 of cognition, even in laboratories. The current developments in artificial intelligence are probably an interesting lever for enhancing the cognition level of the various applications.

**Table 2.** CPPS developments positioning relatively to cognition and development extent

| Cognition level / CPPS Extent | C1 | C2 | C3 | C4 | C5 |
|---|---|---|---|---|---|
| **Industry** | | | [84] | | |
| **Learning Factory** | [70, 76] | | [73] | | |
| **Lab XP** | [74, 79, 81, 82] | [9, 21, 24, 71, 72, 78, 80, 89, 90] | | | |
| **POC** | | [75, 77, 83, 85, 86, 87, 88] | | | |

After several decades of questions about the best network standards to be used in manufacturing context, classical TCP/IP (wireless or not) is now globally used for the high-level communications level. For low-level ones and the interconnection between them, the studied articles rarely give the implementation details. We hope that this framework could encourage the future authors to exhibit their communication infrastructure in order to help the community reaching a consensus. The most promising leads at the time being is probably the introduction of OPC UA or AutomationML [38], which exhibits some service-oriented characteristics that fit well the objectives of initiatives such as Industry 4.0.

What can be finally noticed is the lack of references in Automation and manual modes (see Table 3). The Lab XP for example are generally oriented either on a fully automated system, which seems logical considering the availability of humans for experimentations, but when a human is in the loop, then it is to be in charge of the decisions (Tool mode). Considering the Automation mode, one possibility is that such a mode requires the human to trust the CPPS, which might not be the case at the time being. However, it does not presume of the future applications which might implement massively this mode.

Table 3. CPPS developments positioning relatively to human-machine interaction and development extent

| HMI / CPPS Extent | Manual | Tool | Automation | Full |
|---|---|---|---|---|
| Industry | | [84] | | |
| Learning Factory | [76] | [70] | [73] | |
| Lab XP | | [24, 71, 78, 79, 89, 90] | | [9, 21, 72, 74, 80, 81, 82] |
| POC | | [77, 83, 85, 86, 87, 88] | | [75] |

## 5 Conclusion

This article introduces a new analysis framework for classifying Cyber-Physical Production Systems applications relatively to various items, including their cognitive abilities, their application extent, the interaction with human operators, the distribution of intelligence and the network technologies that are used.

This framework was described and applied to several examples retrieved from literature. From the grid extracted from this analysis, several conclusions are drawn for each encountered application domain, noticing that this framework globally fits the main trends that can be spotted in each specific field.

This framework is meant to be the basis of future classification of future CPPS developments for researchers and practitioners in the field. The objective is to use such a classification in order to ease the relative positioning of those developments and foster their visibility towards a larger audience.

The main perspective about the use of the framework deals with its maintainability, notably considering the technological developments that might occur and modify the perception of the concept. This perspective might be incorporated by the future users of this classification in order to create new levels or even new items so that the framework fits at best the future developments.